# Complex-valued universal linear transformations and image encryption using spatially incoherent diffractive networks

Xilin Yang[1,2,3†]                mikexlyang@ucla.edu

Md Sadman Sakib Rahman[1,2,3†]    mssr@ucla.edu

Bijie Bai[1,2,3]                  baibijie@ucla.edu

Jingxi Li[1,2,3]                  jxlli@ucla.edu

Aydogan Ozcan[1,2,3,*]            ozcan@ucla.edu

[1]Electrical and Computer Engineering Department, University of California, Los Angeles, CA, 90095, USA

[2]Bioengineering Department, University of California, Los Angeles, CA, 90095, USA

[3]California NanoSystems Institute (CNSI), University of California, Los Angeles, CA, 90095, USA

[*]Corresponding author: ozcan@ucla.edu

[†]Equal contribution

# Abstract

As an optical processor, a Diffractive Deep Neural Network (D[2]NN) utilizes engineered diffractive surfaces designed through machine learning to perform all-optical information processing, completing its tasks at the speed of light propagation through thin optical layers. With sufficient degrees-of-freedom, D[2]NNs can perform arbitrary complex-valued linear transformations using spatially coherent light. Similarly, D[2]NNs can also perform arbitrary linear intensity transformations with spatially incoherent illumination; however, under spatially incoherent light, these transformations are non-negative, acting on diffraction-limited optical intensity patterns at the input field-of-view (FOV). Here, we expand the use of spatially incoherent D[2]NNs to complex-valued information processing for executing arbitrary *complex-valued* linear transformations using spatially incoherent light. Through simulations, we show that as the number of optimized diffractive features increases beyond a threshold dictated by the multiplication of the input and output space-bandwidth products, a spatially incoherent diffractive visual processor can approximate any complex-valued linear transformation and be used for all-optical image encryption using incoherent illumination. The findings are important for the all-optical processing of information under natural light using various forms of diffractive surface-based optical processors.

# Introduction

The recent resurgence of analog optical information processing has been spurred by advancements in artificial intelligence (AI), especially deep learning-based methods [1–22]. These advances in data-driven learning methods have also benefitted optical hardware engineering, giving rise to new computing architectures such as Diffractive Deep Neural Networks (D[2]NN), which exploit the passive interaction of light with spatially engineered surfaces to perform visual information processing. D[2]NNs, also referred to as diffractive optical networks, diffractive networks or diffractive processors, have emerged as powerful all-optical processors [9,10] capable of completing various visual computing tasks at the speed of light propagation through thin passive optical devices; examples of such tasks include image classification [11–13], information encryption [14,15], quantitative phase imaging (QPI) [16,17], among others [18–22]. Diffractive optical networks comprise a set of spatially engineered surfaces, the transmission (and/or reflection) profiles of which are optimized using machine learning techniques. After their digital optimization, a one-time effort, these diffractive surfaces are fabricated and assembled in 3D to form an all-optical visual processor, which axially extends at most a few hundred wavelengths ($\lambda$).

Our earlier work [10,23] demonstrated that a spatially coherent D[2]NN can perform arbitrary complex-valued linear transformations between a pair of arbitrary input and output apertures if its design has a sufficient number ($N$) of diffractive features that are optimized, i.e., $N \geq N_i N_o$ where $N_i$ and $N_o$ represent the space-bandwidth product of the input and output apertures, respectively. In other words, $N_i$ and $N_o$ represent the size of the desired complex-valued linear transformation $\boldsymbol{A} \in \mathbb{C}^{N_i \times N_o}$ that can be all-optically performed by an optimized D[2]NN. For a *phase-only* diffractive network, i.e., only the phase profile of each diffractive layer is trainable, the sufficient condition becomes $N \geq 2N_i N_o$ due to the reduced degrees-of-freedom within the diffractive volume. Similar conclusions can be reached for a diffractive network that operates under spatially incoherent illumination: Rahman et al. [24] demonstrated that a diffractive network can be optimized to perform an arbitrary non-negative linear transformation of optical intensity through phase-only diffractive processors with $N \geq 2N_i N_o$. Other optical approaches were also developed to process complex-valued input data with spatially incoherent light [1,25–27]; however, these earlier systems are limited to one-dimensional (1D) optical inputs and do not cover arbitrary input and output apertures, limiting their functionality and processing throughput. An extension of these earlier 1D input approaches introduced the processing of 2D incoherent source-arrays using relatively bulky and demanding optical projection systems that are hard to operate at the diffraction limit of light [28,29].

Here, we demonstrate the processing of complex-valued data with compact diffractive optical networks under spatially incoherent illumination. We show that a spatially incoherent diffractive network that axially spans <100×λ can perform any arbitrary complex-valued linear transformation on complex-valued input data with negligible error if the number of optimizable diffractive features is above a threshold dictated by the multiplication of the input and output space-bandwidth products, determined by both the spatial extent and the pixel size of the input and output apertures. To represent complex-valued spatial information using spatially incoherent illumination, we preprocessed the input information by mapping complex-valued data to a real and non-negative, optical intensity-based representation at the input field-of-view (FOV) of the diffractive network. We term this mapping the 'mosaicing' operation, indicating the utilization of multiple intensity pixels at the input FOV to represent one complex-valued input data point. Similarly, we used a postprocessing step, which involved mapping the output FOV intensity patterns back to the complex number domain, which we termed the 'demosaicing' operation. Through these mosaicing/demosaicing operations, we show that a spatially incoherent D[2]NN can be optimized to perform an arbitrary complex-valued linear transformation between its input and output apertures while providing optical information encryption. The presented spatially incoherent visual information processor, with its universality and thin form factor (<100×λ), shows significant promise for image encryption and computational imaging applications under natural light.

## Results

Figure 1a outlines a spatially incoherent D[2]NN architecture to synthesize an arbitrary complex-valued linear transformation ($A$) such that $o = Ai$, where the input is $i \in \mathbb{C}^{N_i}$, the target is $o \in \mathbb{C}^{N_o}$ and $A \in \mathbb{C}^{N_o \times N_i}$. The mosaicing process involves finding the non-negative (optical intensity-based) representation of each complex-valued element of $i$ using $E$ non-negative values; here, $E$ bases, $e_k$, $k = 0, \cdots, E - 1$ (see Figure 1c), are used for representing the intensity-based encoding of complex numbers. Based on this representation, the 2D input aperture of a spatially incoherent D[2]NN will have $EN_i$ non-negative (optical intensity) values, denoted as $i_r \in \mathbb{R}_+^{EN_i}$, representing the input information under spatially incoherent illumination. The output intensity distribution, denoted with $\widehat{o_r} \in \mathbb{R}_+^{EN_o}$, undergoes a demosaicing process where a complex number is synthesized from the intensity values of $E$ output pixels, yielding the complex output vector $\widehat{o} \in \mathbb{C}^{N_o}$ such that $\widehat{o} \approx Ai$.

In our analyses, we used $E = 3$, except in Supplementary Figure. 2, where $E = 4$ results are shown for comparison. We chose the basis complex numbers as $e_k = \exp\left(jk\frac{2\pi}{E}\right), k = 0, \cdots, E - 1$ such that the set of bases $S$ is closed under multiplication, and the product of any two of the bases in the set is also a basis; for example, for $E = 3$ we have $e_k e_l = e_{(k+l \bmod 3)}$. Based on this representation of information, with $E = 3$ and $e_0, e_1, e_2$, we can decompose any arbitrarily selected complex valued transformation matrix $A$ into $E = 3$ matrices ($A_0, A_1, A_2$) with real non-negative entries such that:

$$A = e_0 A_0 + e_1 A_1 + e_2 A_2 \tag{1}$$

For a given complex-valued input $i = e_0 i_0 + e_1 i_1 + e_2 i_2$, where $i_k \in \mathbb{R}_+$, the corresponding target output vector can be written as:

$$o = Ai = (e_0 A_0 + e_1 A_1 + e_2 A_2)(e_0 i_0 + e_1 i_1 + e_2 i_2) \tag{2}$$

$$o = e_0(A_0 i_0 + A_2 i_1 + A_1 i_2) + e_1(A_1 i_0 + A_0 i_1 + A_2 i_2) + e_2(A_2 i_0 + A_1 i_1 + A_0 i_2) \tag{3}$$

i.e., we have:

$$o_r = \begin{bmatrix} o_0 \\ o_1 \\ o_2 \end{bmatrix} = \begin{bmatrix} A_0 & A_2 & A_1 \\ A_1 & A_0 & A_2 \\ A_2 & A_1 & A_0 \end{bmatrix} \begin{bmatrix} i_0 \\ i_1 \\ i_2 \end{bmatrix} = A_r i_r \tag{4}$$

with a non-negative real-valued matrix $A_r$:

$$A_r = \begin{bmatrix} A_0 & A_2 & A_1 \\ A_1 & A_0 & A_2 \\ A_2 & A_1 & A_0 \end{bmatrix} \tag{5}$$

For $E = 4$, where $e_k e_l = e_{(k+l \bmod 4)}$ and $A = e_0 A_0 + e_1 A_1 + e_2 A_2 + e_3 A_3$, a similar analysis yields:

$$A_r = \begin{bmatrix} A_0 & A_2 & A_3 & A_1 \\ A_2 & A_0 & A_1 & A_3 \\ A_1 & A_3 & A_0 & A_2 \\ A_3 & A_1 & A_2 & A_0 \end{bmatrix} \tag{6}$$

Based on these equations, one can conclude that to all-optically implement an arbitrary complex-valued transformation $o = Ai$ using a spatially incoherent D²NN, the layers of the D²NN need to be optimized to perform an intensity linear transformation $A_r \in \mathbb{R}_+^{E^2 N_i N_o}$ such that $o_r = A_r i_r$. The entire system, upon convergence, performs the predefined complex-valued linear transformation $A$ on any given input data using spatially incoherent light – based on Eqs. 2 and 4. In the following sections, we numerically explore the number of optimizable diffractive features ($N$) needed for accurate approximation of $A$ using a spatially incoherent D²NN.

## Complex-valued linear transformations through spatially incoherent diffractive networks

We numerically demonstrated the capabilities of diffractive optical processors to universally perform any arbitrarily chosen complex-valued linear transformation with spatially incoherent light. Throughout the paper, we used $N_i = N_o = 16$. To visually represent the data, we rearranged the 16-element vectors into $4 \times 4$ arrays of complex numbers, hereafter referred to as the "complex image." We arbitrarily selected a desired complex-valued transformation, $A \in \mathbb{C}^{16 \times 16}$, as shown in Figure 1(b).

To explore the number of diffractive features needed, we trained nine models with varying values of $N$ and evaluated the mean-squared-error (MSE) between the numerically measured ($\widehat{A_r}$) and the target all-optical linear transformation, $A_r$ (see Fig. 2). Our results summarized in Figure 2 highlight that with a sufficient number of optimizable diffractive features, i.e., $N \geq 2E^2 N_i N_o = 2N_{i,r} N_{o,r}$, our system achieves a negligible approximation error with respect to the target $A_r \in \mathbb{R}_+^{48 \times 48}$. In Fig. 2c, we also visualize the resulting all-optical intensity transformation $\widehat{A_r}$ compared to the ground truth $A_r$. In essence, this comparison reveals the spatially varying incoherent point-spread-functions (PSF) of our diffractive system optimized using deep learning; a negligible MSE between $\widehat{A_r}$ and $A_r$ shows that the resulting spatially varying incoherent PSFs match the target set of PSFs dictated by $A_r$.

We also evaluated the numerical accuracy of our complex-valued transformation in an end-to-end manner, as illustrated in Fig. 2d. For this numerical test, we sequentially set each entry of $i$ to $e_0$ and evaluated the corresponding complex output $\hat{o}$ and stacked them to form $\widehat{A_0}$, where the subscript represents that the measurement was evaluated using the complex impulse along the basis $e_0$ as input. Then, we repeated this process for the other two bases to obtain $\widehat{A_1}$ and $\widehat{A_2}$, and stacked these matrices as a block matrix $[\widehat{A_0} | \widehat{A_1} | \widehat{A_2}]$ shown in Figure 2d. Each row of the images $\text{amp}(\hat{o})$ and $\text{phase}(\hat{o})$ in Figure 2d represents one of these complex output vectors, while the corresponding target vectors are presented in the same figure through $\text{amp}(o)$ and $\text{phase}(o)$. The small magnitude of the error $\varepsilon = |\hat{o} - o|^2$ shown in Fig. 2d illustrates the success of this spatially incoherent D²NN model in accurately approximating the complex-valued linear transformation $o = Ai$, implemented for an arbitrarily selected $A$.

## Complex number-based image encryption using spatially incoherent diffractive networks.

In this section, we demonstrate a complex number-based image encryption-decryption scheme using a spatially incoherent D²NN. In the first scheme shown in Figure 1d, the message is encoded into a complex image, employing either amplitude and phase encoding or real and imaginary part encoding. Then, a digital lock encrypts the image by applying a linear transformation ($A^{-1}$) to conceal the original message within the image. At the optical receiver, the encrypted message is deciphered by an optimized incoherent D²NN that all-optically implements the inverse transformation $A$. In an alternative scheme, as depicted in Figure 1e, the key and lock are switched, i.e., the spatially incoherent D²NN is used to encrypt the message with a complex-valued $A$ while the decryption step involves the digital inversion using $A^{-1}$.

For our analysis, we used the letters 'U', 'C', 'L', 'A' as sample messages. 'U' and 'C' are used in amplitude-phase based encoding (Figure 3), whereas 'L' and 'A' are used for real-imaginary based encoding of information (Supplementary Figure S1), forming complex-number-based images. To accurately model the spatially incoherent propagation [24] of light through the D²NN, we averaged the output intensities over a large number $N_\varphi = 20,000$ of randomly generated 2D phase profiles at the input (see Methods section for details).

In Figure 3(a), we show the results corresponding to digital encryption and optical diffractive decryption, i.e., the system shown in Figure 1d. The digitally encrypted complex information $\boldsymbol{i} = \boldsymbol{A}^{-1}\boldsymbol{o}$, together with its intensity representation $\boldsymbol{i_r}$, are shown in Fig. 3(a). The optically decrypted output $\widehat{\boldsymbol{o}}$ (through the spatially incoherent D²NN) and its intensity-based representation $\widehat{\boldsymbol{o_r}}$ are shown in the same figure, together with the resulting error maps, i.e., $|\widehat{\boldsymbol{o}} - \boldsymbol{o}|^2$ and $|\widehat{\boldsymbol{o_r}} - \boldsymbol{o_r}|^2$, which reveal a very small amount of error. This agreement of the recovered and the ground truth messages in both the intensity and complex-valued domains confirms the accuracy of the diffractive decryption process through an optimized spatially incoherent D²NN. Figure 3(b) shows the successful performance of the sister scheme (Figure 1e), which involves diffractive encryption through a spatially incoherent D²NN and digital decryption, also revealing a negligible amount of error in both $\left|\boldsymbol{A}^{-1}\widehat{\boldsymbol{o}} - \boldsymbol{i}\right|^2$ and $|\widehat{\boldsymbol{o_r}} - \boldsymbol{o_r}|^2$. Reported in Supplementary Figure S1, we also conducted a numerical experiment using the letters 'L' and 'A', encoded using the real and imaginary parts of the message. The visualizations are arranged the same way as in Figure 3, where for both schemes depicted in Figure 1d,e, the amount of error between the recovered and the original messages is negligible, affirming the success of using the real and imaginary part-based encoding method.

### Different mosaicing and demosaicing schemes in a spatially incoherent D²NN

How we assign each element in the vector $\boldsymbol{i_r}$ and $\boldsymbol{o_r}$ to the pixels at the input and output FOVs of the diffractive network does not affect the final accuracy of the image/message reconstruction. For example, we can arrange the FOVs in such a manner that the components $\boldsymbol{i_{r,k}}$ corresponding to a basis $e_k$ are assigned to the neighboring pixels, in two adjacent rows, as shown in Supplementary Figure S2a; in an alternative implementation, the assignment/mapping can be completely arbitrary, which is equivalent to applying a random permutation operation on the input and output vectors (see the Methods section). When compared to each other, these two approaches of mosaicing and demosaicing schemes show negligible differences in the error of the final reconstruction of the letters 'U', 'C', 'L', 'A' as shown in Supplementary Figure S2b. These results underscore that the specific arrangement of the mosaicking/demosaicing schemes at the input and output FOVs does not impact the performance of the incoherent D²NN system.

### Discussion

In this manuscript, we employed a data-free PSF-based D²NN optimization method (see the 'Methods' section) [24] since we can determine the non-negative intensity transformation $\boldsymbol{A_r}$ from the target complex-valued transformation $\boldsymbol{A}$ based on the mosaicing and demosaicing schemes; the columns of $\boldsymbol{A_r}$ represent the desired spatially varying PSFs of the D²NN. The advantage of this data-free learning-based

D²NN optimization approach is that computationally demanding simulation of wave propagation with large $N_\varphi$ is not required during the training, i.e., $N_\varphi = 1$ is sufficient for simulating the spatially varying PSFs; hence the training time is much shorter. On the other hand, this approach necessitates prior knowledge of $A_r$, which might not always be available, e.g., for tasks such as data classification. An alternative to this data-free PSF-based optimization approach is to train the diffractive network in an end-to-end manner, using a data-driven direct training approach [24]. This strategy advances by minimizing the differences between the outputs and the targets on a large number of randomly generated examples, thereby learning the spatially varying PSFs *implicitly* from numerous input-target intensity patterns corresponding to the desired task - instead of learning from an explicitly predetermined $A_r$. This direct approach, however, requires a longer training time, necessitating the simulation of incoherent propagation for each training sample on a large dataset.

In our presented approach, the choice of $E$ is not restricted to $E = 3$, as we have used throughout the main text. As another example of encoding, we show the image encryption results with $E = 4$ in Supplementary Fig. S3, where the four bases are $\exp j\frac{\pi}{2}k$ $(k = 0,1,2,3)$. The reconstructed 'U', 'C', 'L', 'A' letters are also reported in the same figure, confirming that given sufficient degrees-of-freedom (with $N \geq 2E^2 N_i N_o$), the linear transformation performances are similar to each other. However, compared to $E = 3$, this choice of $E = 4$ necessitates 4/3 times more pixels on both the diffractive network input and output FOVs – reducing the throughput (or spatial density) of complex-valued linear transformations that can be performed using a spatially incoherent D²NN. Accordingly, more diffractive features and a larger number of independent degrees of freedom (by 16/9-fold) are required within the D²NN volume to achieve an output performance level that is comparable to a design with $E = 3$.

Our framework offers several flexibilities in implementation, which could be useful for different applications. First, the flexibility to arbitrarily permute the input and the output pixels following different mosaicing and demosaicing schemes (as introduced earlier in the Results section) could enhance the security of optical information transmission. A user would not be able to either spam or hack valuable information that is transferred optically without specific knowledge of the mosaicing *and* demosaicing schemes, thus ensuring the security of this scheme. Second, the flexibility in choosing $E$, as discussed above, could be useful in adding an extra layer of security against unauthorized access. Furthermore, we can use different sets of bases for mosaicing and demosaicing operations by applying offset phase angles $\theta_i$ and $\theta_o$, respectively, to the original bases $e_k = \exp\left(jk\frac{2\pi}{E}\right), k = 0, \cdots, E-1$. This will result in a set of modified/encrypted bases: $e_{k,i} = \exp\left(j\left(k\frac{2\pi}{E} + \theta_i\right)\right)$ for mosaicing and $e_{k,o} = \exp\left(j\left(k\frac{2\pi}{E} + \theta_o\right)\right)$ for demosaicing. This powerful flexibility in representation further enhances the security of the system.

Regarding image encryption-related applications, we demonstrated two approaches (Figures 1 d-e) to utilize D²NNs for encryption or decryption. However, it is also possible to deploy a pair of diffractive systems in tandem, with one undertaking the matrix operation $A$ for encryption and the other undertaking the inverse operation $A^{-1}$ for decryption. Furthermore potential extensions of our work could explore a harmonized integration of polarization state controls [30] and wavelength multiplexing [31] to build a multi-faceted, fortified encryption platform. In addition to increasing the data throughput, these additional degrees of freedom enabled by different illumination wavelengths and polarization states would further enhance the security of a diffractive processor-based system.

To conclude, we demonstrated the capability of spatially incoherent diffractive networks to perform arbitrary complex-valued linear transformations. By incorporating various forms of mosaicing and demosaicing operations, we paved the way for a wider array of applications by leveraging incoherent D²NNs for complex-valued data processing. We also showcased potential applications of these spatially incoherent D²NNs for complex number-based image encryption or decryption, highlighting the security benefits arising from the system's flexibility. Our exploration marks a significant stride toward enhanced versatility and robustness in optical information processing with spatially incoherent diffractive systems that can work under natural light.

## Methods

### Linear transformation matrix

In this paper, we use $N_i = N_o = 16$ so that $A \in \mathbb{C}^{16 \times 16}$; see Figure 1b. To generate $A$, we randomly sample the amplitude of each element from the uniform distribution $Uniform(0,1)$ and the phases from $Uniform(0,2\pi)$. For the encryption application, to ensure that the result of inversion is not sensitive to small errors, we performed QR-factorization on $A$ to obtain a condition number of one [32].

### Real-valued non-negative representation of complex numbers

Following Eq. 4, the complex-valued input and target vectors $i \in \mathbb{C}^{N_i}$ and $o \in \mathbb{C}^{N_o}$ are represented by the corresponding real and non-negative intensity vectors $i_r = [i_0^T \cdots i_{E-1}^T]^T \in \mathbb{R}_+^{EN_i}$ and $o_r = [o_0^T \cdots o_{E-1}^T]^T \in \mathbb{R}_+^{EN_o}$, where $i = \sum_{k=0}^{E-1} e_k i_k$ and $o = \sum_{k=0}^{E-1} e_k o_k$. The desired all-optical intensity transformation $A_r$ between $i_r$ and $o_r$ is derived from the target complex-valued linear transformation $A$ following Eqs. 1 and 5. We should note that deriving $A_r$ from $A$ requires mapping each complex element $a$ to its real and non-negative representation $(a_0, \cdots, a_{E-1})$ based on the $E \geq 3$ complex bases $e_k$ such that $a = \sum_{k=0}^{E-1} e_k a_k$. To define a unique mapping, we imposed additional constraints: $a_k = 0$ if $\frac{2\pi}{E} \leq$ phase$(e_k a^*) \leq 2\pi - \frac{2\pi}{E}$, i.e., $a_k = 0$ if the angle between $a$ and $e_k$ is greater than $\frac{2\pi}{E}$; here $a^*$ represents the complex conjugate of $a$. The same constraints were also used while mapping the complex input vectors $i$ to the real and non-negative intensity vectors $i_r$.

### Mosaicing and demosaicing schemes

For mosaicing (demosaicing) assignment of each element of $i_r$ ($o_r$) to one of the $N_{i,r} = EN_i$ ($N_{o,r} = EN_o$) pixels of the 2D input (output), the arrangement of the FOV can be *regular*, e.g., in a row-major order as shown in Supplementary Figure S2 (a), 'Regular mosaicing'. Alternatively, the pixel assignment on the input (output) FOV can follow any arbitrary mapping which can be defined by a permutation matrix $P_i$ ($P_o$) operating on the input (output) vector; see Supplementary Figure S2 (a), 'Arbitrary mosaicing'. For such cases, when ordered in a row-major format, intensities on the input (output) FOVs $i_r$ ($o_r$) can be written as $i_r = P_i[i_0^T \cdots i_{E-1}^T]^T$ ($o_r = P_o[o_0^T \cdots o_{E-1}^T]^T$). Accordingly, such an arbitrary arrangement of pixels was accounted for by redefining the all-optical intensity transformation as $P_o A_r P_i^T$.

## Spatially incoherent light propagation through a D²NN

The 1D vector $i_r$ is rearranged as a 2D distribution of intensity $I(x,y)$ at the input FOV of the D²NN. To numerically model the spatially incoherent propagation of the input intensity distribution $I(x,y)$ through the D²NN, we coherently propagated the optical field $\sqrt{I}\exp(j\varphi)$ through the trainable diffractive surfaces to the output plane, where $\varphi$ is a random 2D phase distribution, i.e., $\varphi(x,y) \sim Uniform(0, 2\pi)$ for all $(x,y)$. If we denote the coherent field propagation operator as $\mathfrak{D}\{\cdot\}$ (see the next subsection), then the instantaneous output intensity is $\left|\mathfrak{D}\{\sqrt{I(x,y)}\exp(j\varphi(x,y))\}\right|^2$ and the time-averaged output intensity $O(x,y)$ for spatially incoherent light can be written as:

$$O(x,y) = \langle \left|\mathfrak{D}\{\sqrt{I(x,y)}\exp(j\varphi(x,y))\}\right|^2 \rangle \qquad (7)$$

The average output intensity can be approximately calculated by repeating the coherent wave propagation $\mathfrak{D}\{\cdot\}$ $N_\varphi$-times, each time with a different random phase distribution $\varphi_r(x,y)$, and averaging the resulting $N_\varphi$ output intensities:

$$O(x,y) = \lim_{N_\varphi \to \infty} \frac{1}{N_\varphi} \sum_{r=1}^{N_\varphi} \left|\mathfrak{D}\{\sqrt{I(x,y)}\exp(j\varphi_r(x,y))\}\right|^2 \qquad (8)$$

We used $N_\varphi = 20{,}000$ for estimating the incoherent output intensity $O(x,y)$ corresponding to any arbitrary input intensity $I(x,y)$. However, for evaluating the spatially incoherent PSFs of the D²NN, $N_\varphi = 1$ is sufficient.

## Coherent propagation of optical fields: $\mathfrak{D}\{\cdot\}$

The propagation of spatially coherent light patterns through a diffractive processor, denoted by $\mathfrak{D}\{\cdot\}$, involves a series of interactions with consecutive diffractive surfaces, interleaved by wave propagation through the free-space separating these surfaces. We assume that these modulations are introduced by *phase-only* diffractive surfaces, i.e., the field amplitude remains unchanged during the light-matter interaction. Specifically, we assume that a diffractive surface alters the incident optical field, symbolized as $u(x,y)$, in a localized manner according to the optimized phase values $\phi_M(x,y)$ of the diffractive features, resulting in the phase-modulated field $u(x,y)\exp(j\varphi_M(x,y))$. The diffractive surfaces are coupled by free-space propagation, allowing the light to travel from one surface to the next. We used the angular spectrum method to simulate the free-space propagation [33]:

$$u(x,y; z = z_0 + d) = \mathcal{F}^{-1}\{\mathcal{F}\{u(x,y; z=z_0)\} \times H(f_x, f_y; d)\} \qquad (9)$$

where $\mathcal{F}\{\cdot\}$ is the two-dimensional Fourier transform and $\mathcal{F}^{-1}\{\cdot\}$ is its inverse operation. $H(f_x, f_y; d)$ is the free-space transfer function corresponding to propagation distance $d$. For wavelength $\lambda$:

$$H(f_x, f_y; d) = \begin{cases} \exp\left(j\frac{2\pi}{\lambda}d\sqrt{1-(\lambda f_x)^2-(\lambda f_y)^2}\right), & f_x^2 + f_y^2 < 1/\lambda^2 \\ 0, & \text{otherwise} \end{cases} \quad (10)$$

The fields were discretized with a lateral sampling interval of $\delta \approx 0.53\lambda$ to accommodate all the propagating modes and sufficiently zero-padded to remove aliasing artifacts[34].

## Diffractive network architecture

We modeled the diffractive surfaces by their laterally discretized heights $h$, which correspond to phase delays $\varphi_M = \frac{2\pi}{\lambda}(n-1)h$, where $n$ is the refractive index of the material. The connectivity between consecutive diffractive layers [9] was kept equal across the diffractive designs with varying $N$ by setting the separation between the layers as $d = \frac{W\delta}{\lambda}$, where the width of each diffractive layer is $W = \sqrt{\frac{N}{K}}\delta$. Here $K$ is the number of diffractive layers; we used $K = 4$ throughout the paper.

## Supplementary Information

- Training and evaluation of spatially incoherent diffractive processors
- Supplementary Figure S1
- Supplementary Figure S2
- Supplementary Figure S3

# Figures and captions

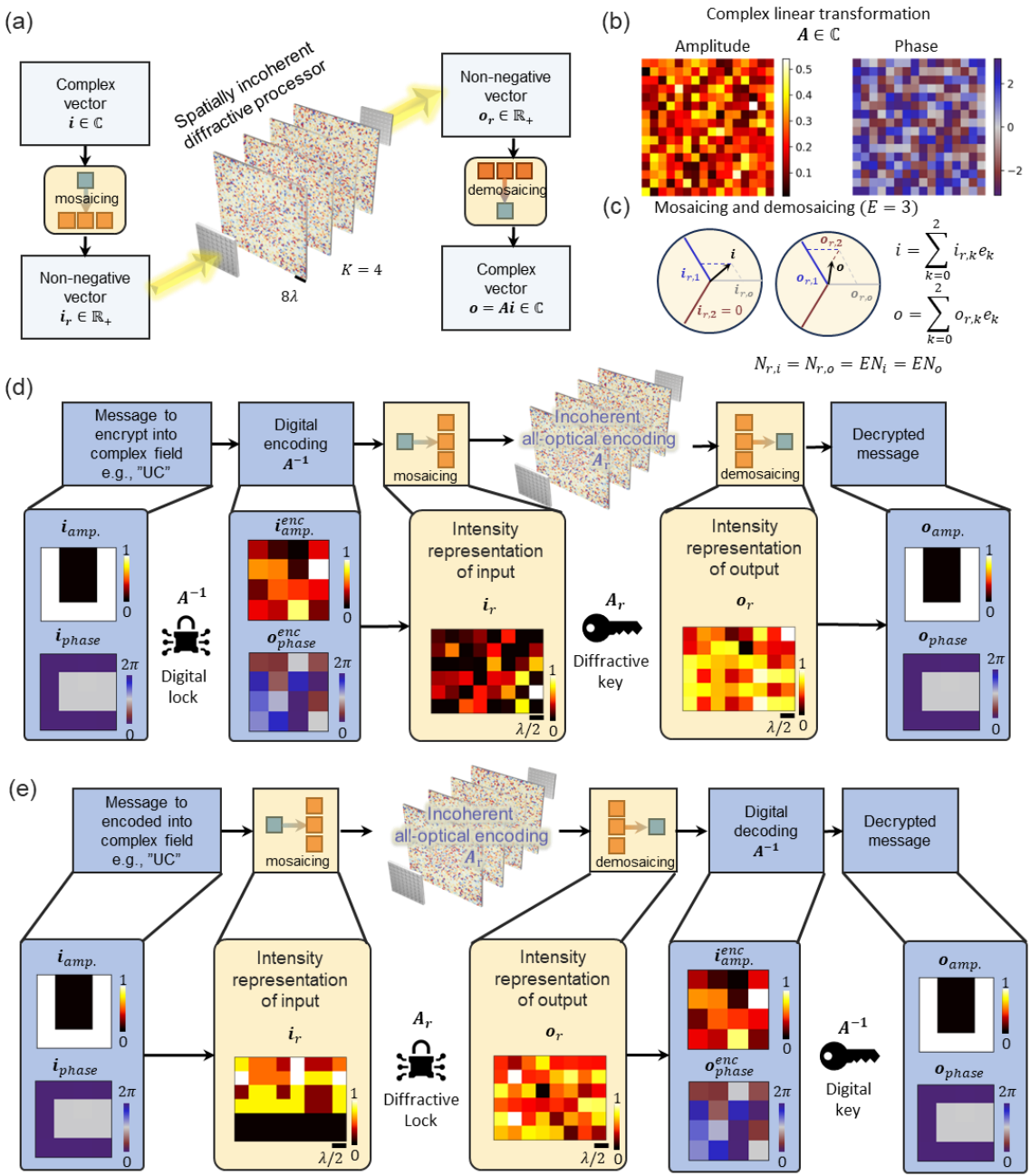

**Figure 1.** (a) Complex-valued universal linear transformations using spatially incoherent diffractive optical networks. (b) Amplitude and phase of the target complex-valued linear transformation. (c) Mosaicing and demosaicing processes. (d-e) Image encryption. In (d), complex-valued images are digitally encrypted ($A^{-1}$), and subsequently decrypted using the diffractive system that performs $A$ (diffractive key). For (e), the encryption is performed through the spatially incoherent diffractive network (diffractive lock) and the decryption is performed digitally (digital key).

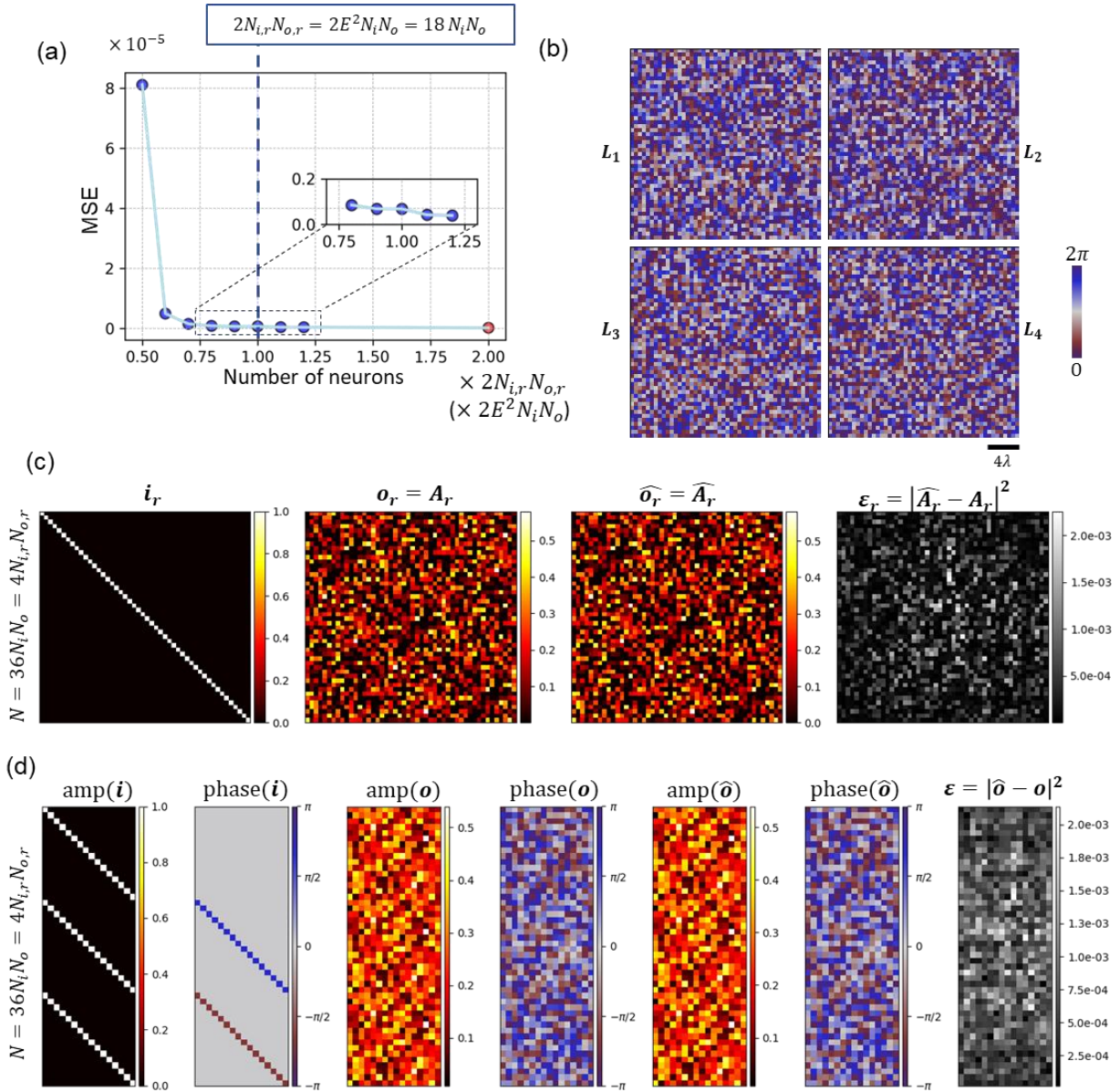

**Figure 2.** Performance of spatially incoherent diffractive networks on arbitrary complex-valued linear transformations. (a) All-optical linear transformation error as a function of the number of diffractive features ($N$). The red dot represents the design corresponding to the results shown in b-d. (b) The phase profiles of the K=4 diffractive layers of the optimized model ($N = 2 \times 2N_{i,r}N_{o,r}$). (c) Evaluation of the resulting all-optical intensity transformation, i.e., the spatially varying PSFs. (d) The complex linear transformation evaluation. For $\varepsilon_r$ and $\varepsilon$, $|\cdot|^2$ represents an elementwise operation.

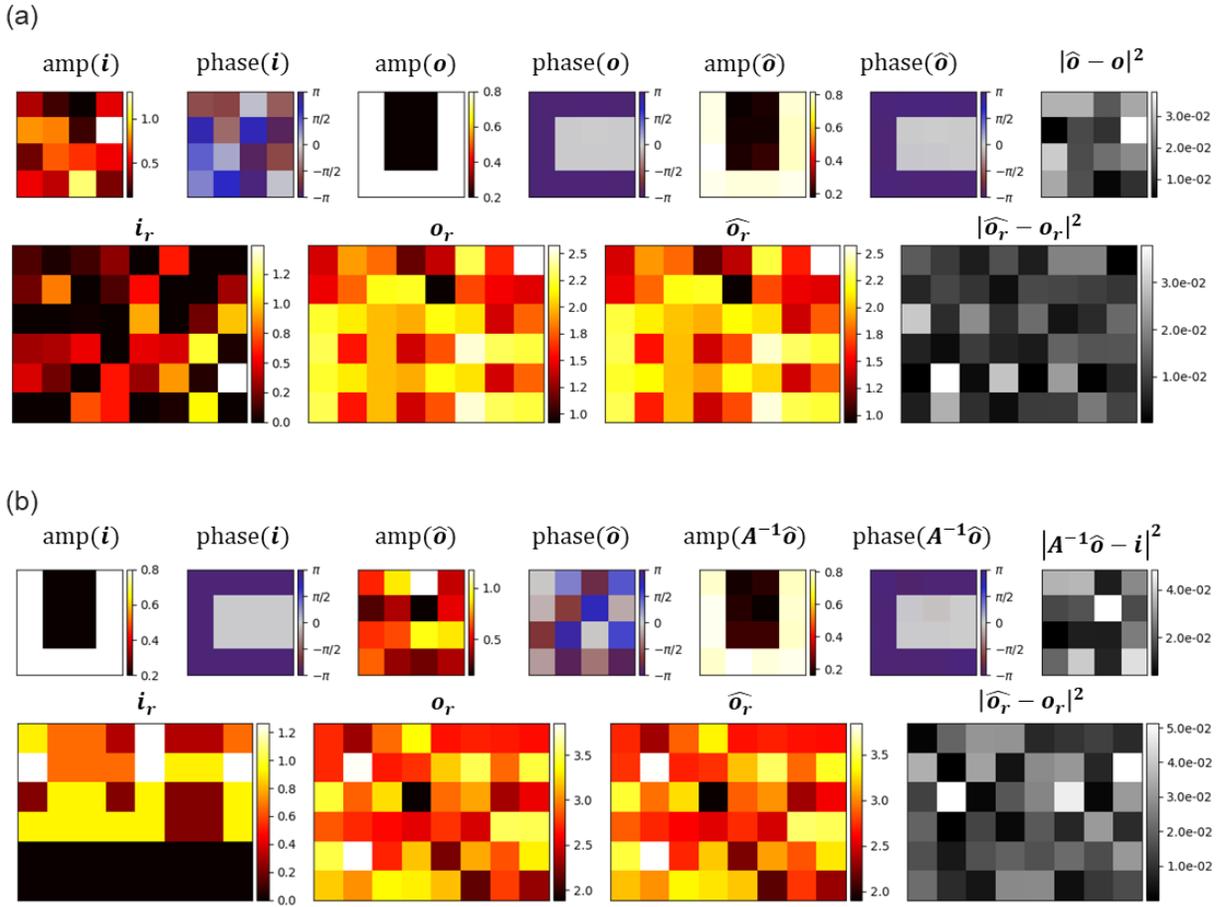

**Figure 3.** Image encryption with the letters 'U' and 'C' encoded into amplitude and phase, respectively, of the complex-valued image. (a) The input, target, output, and the approximation error, both in complex and real non-negative (intensity) domains. The original information is represented by $o$ while $i$ is obtained by digital encrypting $o$ following Figure 1d. (b) The input, output (resulting from optical encryption) and digitally decrypted output and the error between the input and the decrypted output. The result of digital decryption matches the input information. The second row shows the corresponding input, target and output intensities and the approximation error. $|\cdot|^2$ represents an elementwise operation.